# PREPRINT - Tracking the whole-body centre of mass while seated in a wheelchair using motion capture

Félix Chénier[1,2*], Etienne Marquis[1,2], Maude Fleury-Rousseau[1,2]

[1] Mobility and Adaptive Sports Research Lab, Department of physical activity science, Université du Québec à Montréal (UQAM), Montreal, Canada.

[2] Centre for Interdisciplinary Research in Rehabilitation of Greater Montreal (CRIR), Montreal, Canada.

* Corresponding author:

Félix Chénier, PhD
Professor, UQAM | Department of physical activity science
Office SB-4455, Biological Sciences building
141 Président-Kennedy, Montreal QC, H2X 1Y4
chenier.felix@uqam.ca
https://felixchenier.uqam.ca

## Abstract

Estimating the position of the whole-body centre of mass (CoM) based on skin markers and anthropometric tables requires tracking the pelvis and lower body, which is impossible for wheelchair users due to occlusion. In this work, we present a method to track the user's whole-body CoM using visible markers affixed to the user and wheelchair where the user remains seated in their wheelchair, by expressing the pelvis and lower body segments in wheelchair coordinates. The accuracy of this method was evaluated on the anterior-posterior (AP) and medial-lateral (ML) axes by comparing the projected CoM to the centre of pressure measured by four force plates, for 11 able-bodied participants adopting 9 static postures that include extreme reaching postures. The estimation accuracy was within 33 mm (AP) and 9 mm (ML), with a precision within 23 mm (AP) and 12 mm (ML). Tracking the whole-body CoM during wheelchair propulsion will allow researchers to better understand the dynamics of propulsion, which may help devise new approaches to increase the energy transfer from the arms to the ground and reduce the risks of developing musculoskeletal disorders.

## Keywords

Centre of mass, Anthropometry, Wheelchairs, Motion capture

## Introduction

Propelling a wheelchair is a difficult, tiring task, that causes musculoskeletal disorders and pain to the upper body in more than half of wheelchair users (Bayley et al., 1987; Boninger et al., 1997; Curtis and Black, 1999; Samuelsson et al., 2004). The wheelchair is the interface between the user and their environment: any force applied by the user goes through the wheels and wheelchair before reaching the ground. Therefore, pursuing research in improving the propulsion efficiency and minimizing the risk of disorders requires a good understanding of the dynamics of the wheelchair.

Wheelchair dynamics are often modelled using a simple mass and resistance model (van der Woude et al., 2001). However, because the rolling resistance coefficient of the front wheels is generally higher than that of the rear wheels, the user must overcome increasing resistance as the push progresses from the beginning to the end, because their weight is increasingly distributed toward the front wheels. Bascou et al. (2012) found that a variation of weight distribution from 29% to 64% on the caster wheels induces a change in total rolling resistance of 52%, for a same total weight.

In addition, within a push cycle, the user's centre of mass (CoM) accelerates and decelerates relative to the wheelchair, which creates an inertial force that accelerates the wheelchair during the first part of the recovery phase and decelerates the wheelchair during the first part of the push phase (Sauret et al., 2013, 2009). Modelling these inertial forces was shown to better predict wheelchair speed than based on propulsion moments alone, in a study with 19 experienced manual wheelchair users who propelled a wheelchair on the ground (Chénier et al., 2016). This phenomenon has also been observed in wheelchair racing (Moss et al., 2005; Poulet et al., 2022).

To better understand the dynamics of manual wheelchair propulsion, we must model the user and wheelchair as a multi-body system, or minimally estimate the trajectory of the user's whole body CoM during the propulsion. A common method to estimate the CoM is to measure the movement of the segments and find their centres of mass statistically using anthropometric tables. These tables are generally obtained using regression equations from anthropometric data based on cadaveric studies, photogrammetry or medical imaging (Dumas and Wojtusch, 2018). Dumas et al. (2007) adjusted data from McConville et al. (1980) and Young et al. (1983) so that they can be represented in the segments' general coordinates, following the recommendations of the International Society of Biomechanics (ISB) (Wu et al., 2005, 2002). They also provided regression equations to reconstruct joint centres such as the lumbar cervical joint centres based on bony landmarks. However, these equations could not be used for wheelchair-bound people, because the pelvis is occluded by the backrest and wheels. While simplified marker sets have been developed to reduce the number of required markers (Tisserand et al., 2016), they still need to track the pelvis. It is therefore impossible to track a wheelchair user's movement in real, dynamic propulsion conditions in a way that would allow calculating the trajectory of their CoM.

In this paper, we present a method to estimate the position of the user's whole-body CoM in wheelchair-bound people, using a 15-segment multi-body model (head, thorax, pelvis, 2 upper arms, 2 forearms, 2 hands, 2 thighs, 2 shanks, 2 feet). The pelvis and lower bodies are considered joined to the wheelchair and therefore do not need to be tracked during propulsion. Only the head, thorax, upper arms, forearms, hands, and wheelchair need to be tracked. We assessed the precision and accuracy of this model for eleven participants who adopted different seated positions in a manual wheelchair.

# Methods

## Participants

Eleven able-bodied participants (five men, six women) were recruited, with the only inclusion criteria being to be adult. The protocol has been approved by the Institutional Committee for Research with Humans (CIÉREH), certificate 2879_e_2018. Participants were aged 26.5 ± 5.7 years, had a weight of 73.8 ± 12.7 kg, a height of 1.71 ± 0.12 m, and a body-mass index of 25.0 ± 2.5 kg/m². Nine were right-handed and two were left-handed.

## Instrumentation

Unique rigid clusters of five reflective markers were affixed bilaterally on the participants' upper arms (near the elbow), forearms (near the wrist), thorax and head, using standard GoPro accessories. Reflective markers were also affixed bilaterally on the 2nd and 5th metacarpal heads (in white in Fig. 1). We assumed that the pelvis would not slide on the seat, and that the possible small movements of the user's lower body and pelvis would not significantly alter the CoM. Therefore, instead of tracking the lower body and pelvis individually, we considered these segments as rigidly tied to the wheelchair.

All participants used the same wheelchair (Invacare 9000) with four reflective markers installed on the frame to create a rigid wheelchair cluster (in white in Fig. 1). The centres of the rear wheels and the four wheel-ground contact points were probed and referenced to this wheelchair cluster, with the front wheels trailing backward as pictured in Fig. 1 and 2. The tires were fully inflated to minimize the wheel-ground contact size. The marker positions were recorded at a sampling frequency of 120 Hz using a 10-camera Optitrack system.

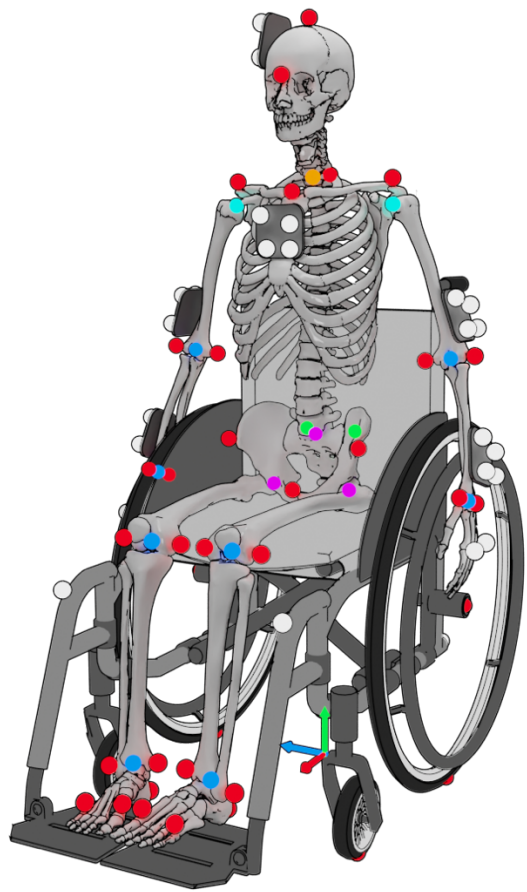

Figure 1. Sequence of marker reconstruction from rigid clusters to joint centres. Red: Probed points. Green: Posterior-superior iliac spines. Magenta: Lumbar and hip joint centres. Orange: Cervical joint centres. Cyan: Shoulder joint centres. Blue: Elbow, wrist, knee and ankle joint centres.

To evaluate the accuracy and precision of the CoM estimation method, the centre of pressure (CoP) was measured using four individual force plates (AMTI OR6-7-2000) synchronized with the motion capture system, each force plate measuring the vertical reaction force under one wheel at a sampling frequency of 1000 Hz (Fig. 2).

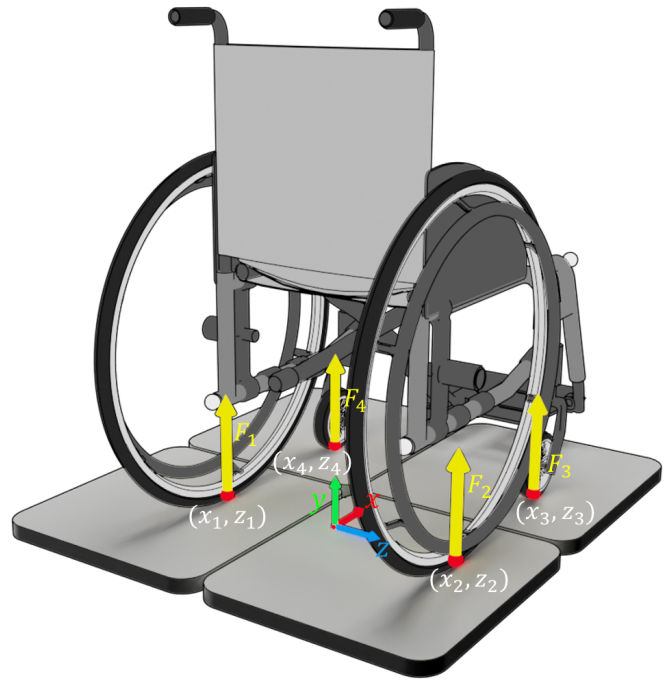

Figure 2. Calculation of the centre of pressure using four vertical forces.

## Calibration

After zeroing the force plates with the empty wheelchair being on the force plates, the participants were asked to take place into the wheelchair and to sit on the forepart of the seat, so that their back did not touch the backrest. This first calibration step, used to reveal the posterior-superior iliac spines, was the only one that required a displacement of the participant in their wheelchair. In this steady position, these five points were probed to create a 5-point pelvis point cloud:

- Left/right anterior-superior iliac spines (LASIS, RASIS)
- Left/right posterior-superior iliac spines (LPSIS, RPSIS)
- Pubic symphysis (SYM)

Participants were then asked to sit with their back leaning against the backrest, in a standard, comfortable position. Their legs and thighs were strapped to the wheelchair to limit their movement. In this position, the following markers were probed and added to their respective clusters:

- 11 points added to the wheelchair cluster:
    - Left/right anterior-superior iliac spines (LASIS, RASIS)
    - Pubic symphysis (SYM)
    - Left/right, lateral/medial femoral epicondyles

- o Left/right, lateral/medial malleoli
- 4 points added to the thorax cluster:
  - o 7th cervical
  - o Left/right acromions
  - o Incisura jugularis (suprasternal notch)
- 2 points added to the head cluster:
  - o Head vertex
  - o Sellion
- 4 points added to the left/right upper arm clusters:
  - o Left/right, lateral/medial humeral epicondyles
- 4 points added to the left/right forearm clusters:
  - o Left/right, ulnar/radial styloids

Finally, a quick static acquisition of two seconds was performed in a neutral position, as in Fig. 1, with a straight thorax, the arms straight down, and the palms against the wheels. The following method was used to generate each point of Fig. 1 during this 2-second static acquisition:

**1. Probed points:** Every probed points (in red in Fig. 1) were reconstructed by tracking the 7 rigid clusters.

**2. Posterior iliac spines:** Both LPSIS and RPSIS (in green in Fig. 1) were added to the wheelchair cluster by registering the LPSIS, RPSIS, LASIS, RASIS, SYM point cloud (as probed with the participant being on the fore-part of the seat) to LASIS, RASIS and SYM (as probed in the final seated position).

**3. Lumbar and hip joint centres**: The position of the lumbar joint centre and hip joint centres were calculated based on Dumas et al. (2007), Reynolds et al. (1982) and Reed et al. (1999) using LASIS, RASIS and SYM. These three new points (in magenta in Fig. 1) were added to the wheelchair cluster.

**4. Cervical joint centres:** The position of the cervical joint centre was calculated based on Dumas et al. (2007) and Reed et al. (1999) using C7, the suprasternal notch, and the lumbar joint centre, the latter being used to complete the sagittal plane. This new point, shown in orange in Fig. 1, was added to the thorax cluster.

**5. Shoulder joint centres:** The position of both shoulder joint centres was estimated at 17% of the inter-acromial distance, directly below the acromion, based on Michaud et al. (2016) and Rab et al. (2002). These two new points (in cyan in Fig. 1), were added to the respective upper arm clusters.

**6. Extremities joint centres:** The joint centres of the elbows, wrists, knees and ankles are the middle of the joint's two probed landmarks (in blue in Fig. 1):

- Elbow joint centres: midpoint of lateral/medial humeral epicondyles, in their respective upper arm cluster.
- Wrist joint centres: midpoint of ulnar/radial styloid processes, in their respective forearm clusters.
- Knee joint centres: midpoint of lateral/medial femoral epicondyles, in the wheelchair cluster.
- Ankle joint centres: midpoint of lateral/medial malleoli, in the wheelchair cluster.

Based on these extended clusters, each point of Fig. 1 can be reconstructed in any posture: the upper body points using their respective clusters, and the pelvis and lower body points using the four wheelchair markers.

## Tasks

The participants were asked to adopt different static postures as shown in Fig. 3, each lasting for 2 seconds. These postures were selected to maximize the shift of centre of mass on both anterior-posterior and medial-lateral axes. Each posture was assessed three times, and the same set of recordings were used for the neutral posture in both Fig. 3a and Fig. 3b, for a total of 9×3 = 27 acquisitions. The participants were asked to move as little as possible during the recordings and were allowed to grab the pushrims during the left and right reach postures to stabilize their body. Wheelchair brakes were applied all the time. To limit possible interactions between successive postures (e.g., a participant who would keep their trunk laterally inclined for all trials of antero-posterior axis), participants were asked to return to a neutral position between each acquisition, and the order of the 27 acquisitions was randomized. The position of the rigid marker clusters and the ground reaction forces were recorded for each acquisition.

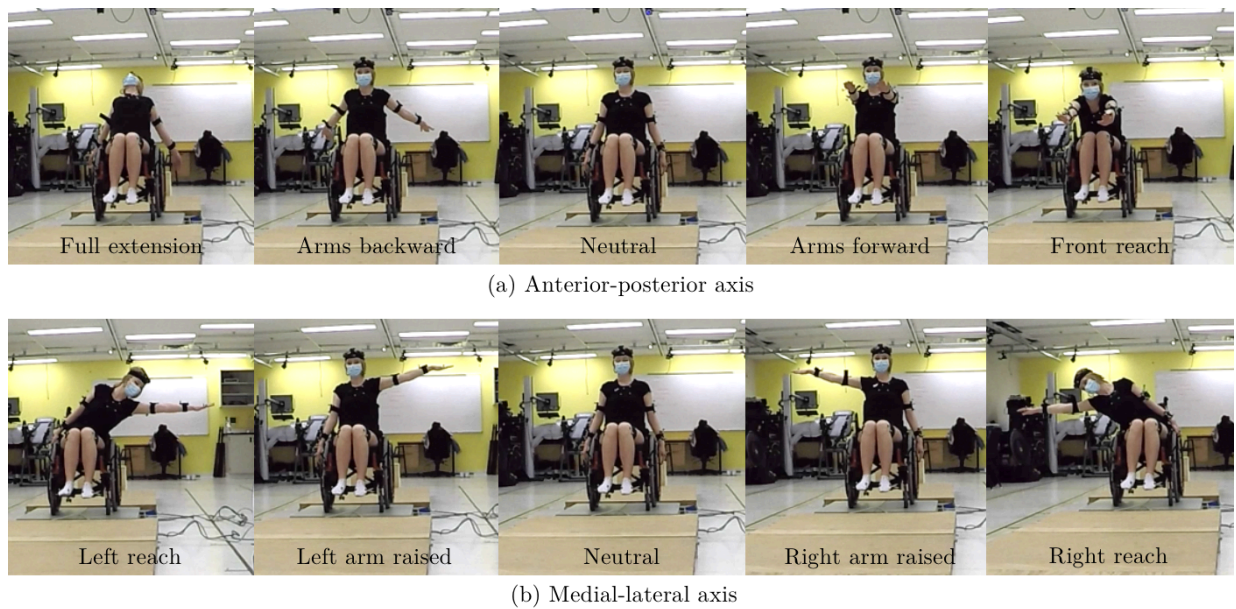

Figure 3. The different postures used to validate the estimation method.

## Data processing

**CoM estimation using motion capture:** Every marker of Fig. 1 was reconstructed by tracking the rigid clusters. The segments LCS were then defined using the joint centres and body landmarks following the recommendations of the ISB (Wu et al., 2005, 2002). The mass and CoM position of each segment were calculated locally in the segments LCS, using values from Dumas et al. (2007), based on anthropometric data from McConville et al. (1980) and Young et al. (1983). The global CoM was finally calculated by weighting these individual centres of mass and was expressed in the wheelchair LCS, defined as z pointing right (using the rear wheels' centres), y being upward, x being forward, and the origin being on the ground at the middle point between the rear wheels as illustrated in Fig. 1 and 2.

**CoP measurement using the force plates:** The reference CoP was measured using the vertical components of the four ground reaction forces (yellow in Fig. 2). Since the force plates had been zeroed with the empty wheelchair, these forces represent the CoP of the participants alone, excluding the wheelchair. The four points of force application, which are the four wheel-ground contact points (red in Fig. 2), had been probed previously in reference to the wheelchair, which allowed calculating the reference CoP directly in the wheelchair LCS in both anterior-posterior ($CoP_{AP}$) and medial-lateral ($CoP_{ML}$) components:

$$CoP_{AP} = \frac{\sum_{i=1}^{4} F_i x_i}{\sum_{i=1}^{4} F_i}$$

$$CoP_{ML} = \frac{\sum_{i=1}^{4} F_i z_i}{\sum_{i=1}^{4} F_i}$$

For each acquisition, both the estimated CoM and the reference CoP were averaged over the 2-second length of the acquisition. The projection of the estimated CoM on the ground, which is coincident to the CoP in static acquisitions, was compared to the reference CoP measured using the force plates.

### Outcome measures

Accuracy was defined as the mean difference between the estimated and reference CoP, over all acquisitions of a given posture. Precision was defined as the standard deviation of the difference between the estimated and reference CoP, over all acquisitions of a given posture.

## Results

The accuracy and precision are shown for each posture in Table 1. The estimation accuracy was generally biased posteriorly, with an accuracy of -33 mm in full extension, to +2 mm in front reach. The precision error was generally below 15 mm, with a notable exception of 23 mm in front reach.

Table 1. Accuracy and precision of the CoP estimation (mm)

| Posture | AP Accuracy | AP Precision | ML Accuracy | ML Precision |
|---|---|---|---|---|
| Full extension | -33 | 10 | 5 | 5 |
| Arms backward | -22 | 13 | 5 | 5 |
| Neutral | -25 | 13 | 5 | 4 |
| Arms forward | -20 | 15 | 5 | 4 |
| Front reach | 2 | 23 | 4 | 7 |
| Left reach | -18 | 12 | 5 | 12 |
| Left arm raised | -21 | 12 | 9 | 5 |
| Right arm raised | -25 | 13 | -1 | 5 |
| Right reach | -17 | 13 | 6 | 7 |

The differences between the estimated and reference values are shown separately for the AP and ML axes as two Bland-Altman plots in Fig. 4. On the AP axis, the mean estimation error was -20 mm, with a 95% confidence interval of -56 mm to +19 mm. A slight correlation may exist between the difference and mean value, with a Pearson correlation coefficient of ρ = 0.59. On the ML axis, the mean estimation error was 5 mm, with a 95% confidence interval of -10 mm to +20 mm. No correlation was observed between the difference and mean value (ρ = -0.03).

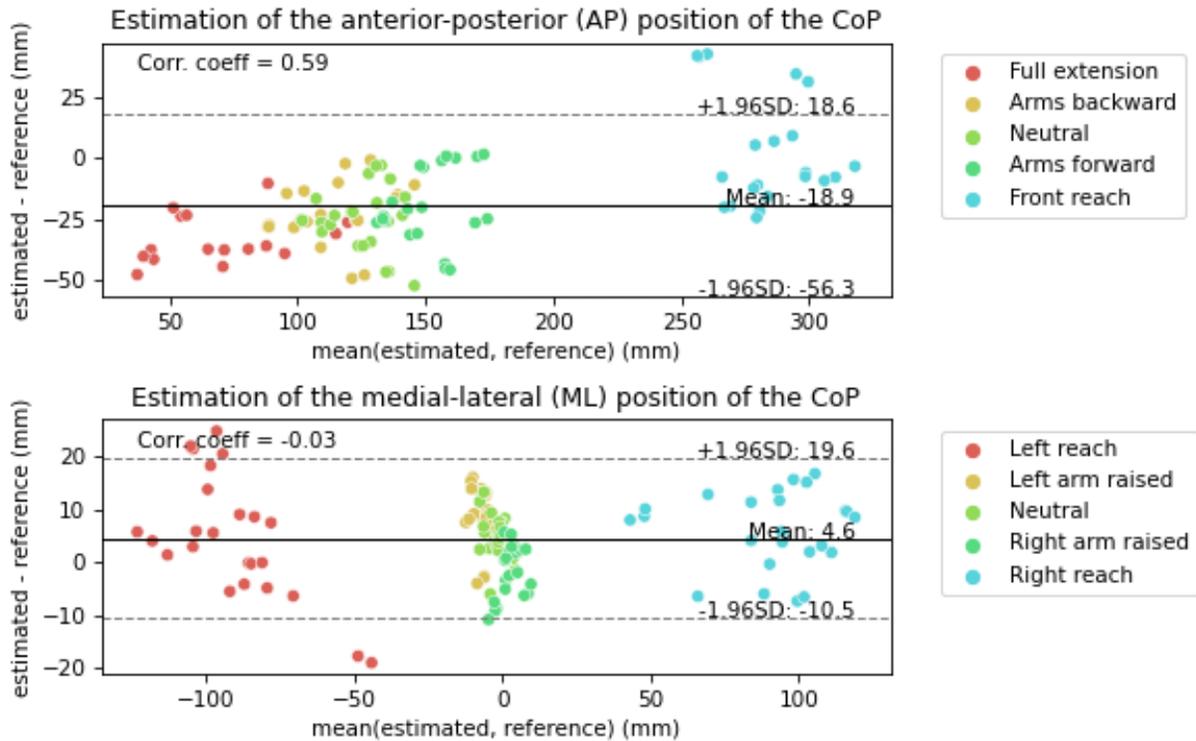

Figure 4. Centre of pressure (CoP) estimation as a function of the posture.

# Discussion

This study aimed to present and test a method to track the whole-body CoM of a seated person in a wheelchair, using only rigid marker clusters affixed to the upper body, thorax, head, and wheelchair. While there was generally a posterior bias in the AP estimation, its precision was generally contained within 5 to 15 mm. This is, to our knowledge, the first verified method to calculate the CoM of wheelchair-bound users using marker tracking. Another method by Kollia et al. (2012) used a volumetric model to estimate the CoM of a seated person based on pairs of photographs; however, this method is also limited by occlusion from the wheelchair, and could hardly be used for CoM tracking in real propulsion conditions.

During the experiments, we realized that the participants did not always keep their trunk straight, and their pelvis rotation was not always constant relative to the wheelchair. For instance, in the full extension posture, the lumbar portion or their trunk was blocked by the backrest, which made them compensate by hyperextending their trunk. This may have contributed to the negative AP estimation error of -33 mm since our model wrongly estimated the trunk as a straight line that would even pass through the backrest (Fig. 5b). In the front reach posture (Fig. 5a), their pelvis was inclined forward, which would move the real CoM forward; however, their trunk was also flexed, which would move the real CoM backward. These two errors seemed to have compensated each other, which may explain the low AP estimation error of 2 mm in this posture. Although measuring the real pelvis incline remains a challenge because of potential occlusion by the abdomen, these results suggest that separating the trunk into two segments could generally improve the CoM estimation. Dumas et al. (2015) published an updated method and inertial values to separate the trunk into abdominal and thoracic components. This method requires three additional points (spine and left/right ribs at 10th thoracic spine) that were not recorded in our work, but that should be relatively easy to probe and reconstruct in reference to the thorax rigid cluster. These three points should be included in future studies to assess if this separation could reduce the trend in AP estimation error. Contrarily to the AP axis, no correlation was observed between the posture and error on the ML axis: the spine is indeed much more mobile in flexion/extension than in lateral deviation, and therefore combining the abdominal and thoracic components into one straight line, as we did, may be more valid for ML movements.

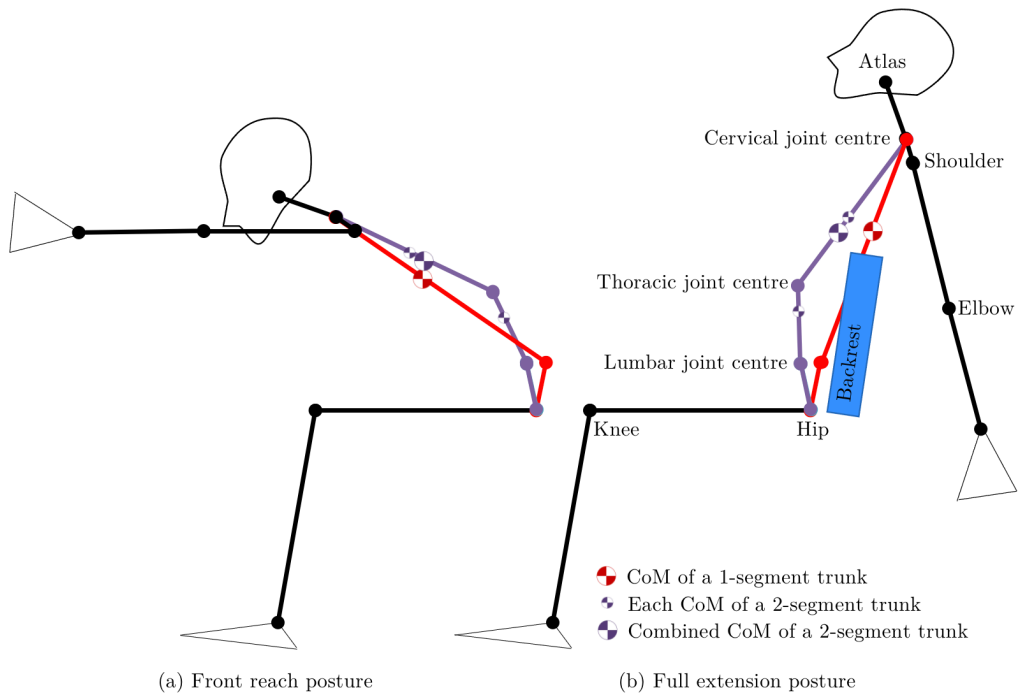

Figure 5. Effect of trunk segmentation on CoM estimation.

Among other studies that measured the accuracy of CoM estimation based on kinematics, Pavei et al. (2017) tested different kinematic models where one participant walked, fast walked, ran and skipped on an instrumented treadmill with motion capture. Their best model was a 14-segment model based on inertial properties from De Leva et al. (1996), with a measured 3D root-mean-square error of 3 to 5 mm. This is significantly lower than ours. In a study with similar instrumentation and validation method where 3 men performed squats, fast squats, lower limb lifting, and upper limb lifting on a force plate, Mapelli et al. (2014) estimated the CoM using a multi-body kinematic model of 10 bodies (no hands, no feet, and one combined pelvis/trunk segment). They obtained a root-mean-square error on the ML axis of 14.0 ± 13.0 mm, which is twice to three times higher than our error figures. They, however, obtained a lower error than we did for the AP axis, with 17.5 ± 6.2 mm. The multi-body model and body-segment inertial properties used in Pavei et al. (2017) may have been more representative of the participants than those used in Mapelli et al. (2014).

Note that both these studies used a double integration of the ground-reaction forces as a reference for their CoM estimation. While this is often considered as a gold standard to measure the CoM displacement (Mapelli et al., 2014; Minetti et al., 2011; Pavei et al., 2017), it assesses only the variation of the CoM position during a task (e.g., one gait cycle) because initial conditions of CoM position and velocity are required as integration constants. Therefore, this method could not measure the absolute CoM estimation error, only the accuracy of the CoM excursion. This is the main reason that made us choose the CoP as a reference, along with the requirement that the whole body is on an instrumented surface during an appreciable duration (which is impossible with the large footprint of a wheelchair). In another study where, as in ours, the CoP was used as a reference, Chen et al. (2011) compared three models for 20 able-bodied individuals who stood in different positions on a force plate, and obtained errors of 11 to 25 mm, which is similar to our results. This is encouraging because our postures were designed to maximize the CoM range, which normally generates larger error, compared to these other studies where the CoM range was limited by the standing postures adopted by the participants.

Among the limitations of our study, the reconstruction of LPSIS and RPSIS based on a five-point cloud (calibration step 2) has not been done before and may be sensitive to error for a number of reasons: (1) the participants need to stay immobile during the probing; (2) three of the five pelvis points need to be probed twice (LASIS, RASIS, SYM) and these points are difficult to probe because of abdominal flesh and clothes. This may create a misalignment between the reconstructed and real pelvis. A new study that focuses only on pelvis reconstruction will be needed to assess this error specifically. As two other limitations, the body segment inertial properties were not personalized to the users but instead taken directly from anthropometric tables, and homogeneous able-bodied participants were recruited. Both limitations combined are expected to generate higher errors in wheelchair users than those measured in our study, mostly in the most extreme postures, because weight distribution is usually different in wheelchair users than able-bodied people. Nonetheless, the postures we assessed are more extreme than those adopted during the propulsion of a wheelchair. In any case, a logical follow-up for this work would be to personalize the body-segment inertial properties, and a starting point could be to calculate new inertial properties by using a volumetric model as described in the aforementioned work from Kollia et al. (2012).

# Conclusion

We presented a method to track the whole-body CoM position of a person seated in a wheelchair, by tracking rigid marker clusters affixed to the wheelchair and body. The novelty of this method is that instead of tracking the pelvis and lower body, which is impossible with wheelchair users, it reconstructs these segments in the wheelchair LCS. Compared to a reference method that measured the CoP using force platforms, the estimation accuracy ranged from -33 to 2 mm in AP direction and from -1 to 9 mm in ML direction, with a precision of 5 to 23 mm. These results are comparable to other estimation methods in standing tasks but are only representative of able-bodied individuals. We expect that a personalization of body-segment inertial properties will be needed for wheelchair users. Tracking the whole-body CoM during the propulsion of a wheelchair will help researchers to better understand the dynamics of wheelchair propulsion, and to devise adaptations to the wheelchair, user interface, or user technique, to enhance their propulsion efficiency and reduce the risk of upper body disorders.

# Acknowledgments

We would like to thank Hervé Lasbats for his participation in data collection, and Nicolas Fleury-Rousseau for his participation in data processing. This work was funded by the Natural Sciences and Engineering Research Council of Canada (NSERC) for Maude Fleury-Rousseau's funding, and the Fonds de recherche du Québec—Nature et technologies (FRQNT) for Etienne Marquis' funding.

# Conflict of interest statement

We hereby declare that this work is free from any real or apparent conflict of interest.